\documentclass[aps, prl, two column, amssymb, amsmath, showpacs, superscriptaddress]{revtex4-1}
\usepackage[T1]{fontenc}
\usepackage[latin9]{inputenc}
\usepackage{amsmath}
\usepackage{graphicx}
\usepackage{amssymb}
\usepackage{esint}
\usepackage{SIunits}
\usepackage{float}

\def\be{\begin{equation}}       \def\ee{\end{equation}}
\def\bea{\begin{eqnarray}}      \def\eea{\end{eqnarray}}
\def\ba{\begin{array} }
\def\ea{\end{array} }

\def\=>{\Rightarrow}
\def\>{\rightarrow}

\def\eye2{Fathbb{I}}

\def\Eq#1{Eq.~(\ref{#1})}

\renewcommand{\v}[1]{{\bf #1}}

\newcommand{\s}{{\sigma}}

\renewcommand{\>}{\rangle}

\newcommand{\e}{\epsilon}

\begin{document}

\title{Time-reversal-invariant topological superconductivity in n-doped BiH}

\author{Fan Yang}
\thanks{\texttt{yangfan\_blg@bit.edu.cn}}
\affiliation{School of Physics, Beijing Institute of Technology, Beijing, 100081, China}
\author{Cheng-Cheng Liu}
\affiliation{School of Physics, Beijing Institute of Technology, Beijing, 100081, China}
\author{Yu-Zhong Zhang}
\affiliation{Shanghai Key Laboratory of Special Artificial Microstructure Materials and Technology, \\
School of Physics Science and Engineering, Tongji University, Shanghai 200092, China}
\author{Yugui Yao}
\thanks{\texttt{ygyao@bit.edu.cn}}
\affiliation{School of Physics, Beijing Institute of Technology, Beijing, 100081, China}
\author{Dung-Hai Lee}
\affiliation{
Department of Physics,University of California at Berkeley,
Berkeley, CA 94720, USA}
\affiliation{Materials Sciences Division,
Lawrence Berkeley National Laboratory, Berkeley, CA 94720, USA}
\begin{abstract}
Despite intense interest and considerable works, definitive experimental evidence for time reversal invariant topological superconductivity is still lacking. Hence searching for such superconductivity in real materials remains one of the main challenges in the field of topological material. Previously it has been shown that in the buckled honeycomb lattice structure, hydrogenated single bilayer Bi, namely BiH, is a topological insulator. Here we predict that upon n-type doping, BiH is a time reversal invariant topological superconductor. Interestingly the edge states of such superconductor consists of both helical complex fermion modes and helical Majorana fermion modes. 

\end{abstract}
\maketitle

\section{I. Introduction}
Intrinsic and symmetry protected topological states have attracted great interest in condensed matter physics recently\cite{XiaogangBook, Xiechen}. In particular, symmetry protected free fermion topological phases have been intensively studied both theoretically and experimentally\cite{KaneRMP, XiaoliangRMP}. For example 2D and 3D topological insulators protected by the time-reversal (TR) symmetry have been theoretically predicted\cite{KaneMele1, KaneMele2, Shoucheng1, FuKane, Moore, Roy1} and experimentally verified\cite{Molenkamp,Hasan}. In addition, possible free fermion topological insulators and superconductors in various spatial dimension have been classified\cite{Schnyder, Kitaev1}. Proposals for realization of time reversal invariant (TRI) superconductivity include the proximity induction\cite{prox1,prox2} and new superconducting materials\cite{Liangfu1,Jingwang,Wan}. However despite focused theoretical and experimental efforts\cite{Roy2,Schnyder,Kitaev1,XiaoliangTRI,Zhangfan,Liangfu1,Liangfu2,Jingwang,Wan,Brydon,Hong,Qianghua,BiSeCu,debate1,debate2,debate3}, there is no consensus on the presence of TRI topological superconductivity in any known system yet. For example whether the superconductivity in Cu$_{x}$Bi$_{2}$Se$_3$\cite{BiSeCu} is a topological one\cite{Liangfu1,Liangfu2} is very much under debate\cite{debate1,debate2,debate3,debate4}.

In this paper we present the theoretical evidence that the n-type doped single bilayer BiH is a TRI topological superconductor. The starting point of our analysis is a tight-binding model of the band structure of BiH. The tight binding parameters are chosen to reproduce the band dispersion of earlier first principle calculations\cite{Ours1,Ours2}. In particular, in the absence of doping, it yields a QSH insulator with a large indirect band gap. We model the electron correlation by the intra and inter-orbital Coulomb repulsion $U,V$ and the Hund's rule coupling $J_H$. These parameters are obtained from constraint density functional theory calculations. Upon electron doping, we perform a mean-field pairing instability analysis which predicts the dominant pairing channel as $(p+ip^{\prime})_{\uparrow\uparrow}, (p-ip^{\prime})_{\downarrow\downarrow}$ and hence leads to a TRI topological superconducting state. We stress that although the analysis in this paper is done for BiH, we believe similar physics should hold in other bilayer Bi-Hydride/Halide materials.
\begin{figure*}
\includegraphics[width=7in]{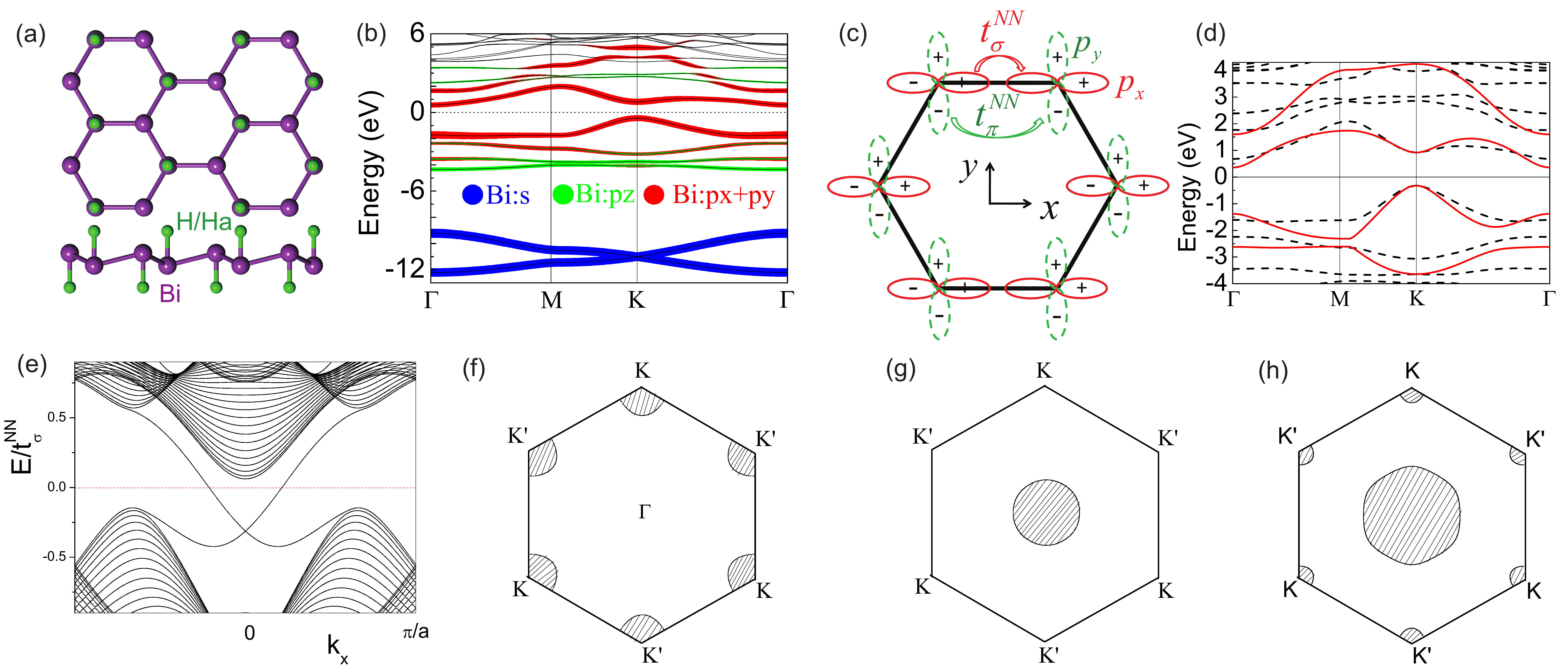}
\caption{(color online).(a) The structure of hydrogenated single bilayer Bismuth, BiH. Upper panel: top view and lower panel: side view. The Bi atoms in each layer form a sublattice of the buckled honeycomb lattice. The Bi-atoms in the upper/lower layer are bonded by hydrogens from above/below. This structure is also applicable for the Bismuth Halide compounds. (b) The band structure of the BiH single bilayer. The size of the symbols is proportional to the weight of the band eigenfunctions on different atomic orbitals (color coded). For undoped compound the Fermi level is set to zero. (c) The arrangement of the $p_x$ and $p_y$ orbitals in a hexagonal plaquette, and the $\s$ and $\pi$ bonds between them. (d) A comparison of the band structures in the presence of SOC for BiH calculated using the first principle (black dotted curve) and the tight binding methods (\ref{eq:H0})(red solid curve). The tight binding parameters are given in Appendix A. (e) The edge spectrum (for the``zigzag" edges) computed using the tight-binding model. (f)-(h)  The Fermi pockets for 5\% hole doping (f) and 4\%, 10\% ((g) and (h)) electron doping respectively.}\label{fig:geomerybandFS}
\end{figure*}

\section{II. Material and Model}
\subsection{A. The crystal and band structures}

 A schematic representation of the optimized crystal structure for the bilayer BiH is shown in Fig.\ref{fig:geomerybandFS}(a). The bilayer of Bi atoms form a buckled honeycomb lattice, with  sublattice A and B hydrogenated from above and below, respectively. The point group is $D_{3d}$ which possesses the inversion symmetry. The band structure is shown in Fig.\ref{fig:geomerybandFS}(b), from which one finds that the main component near the Fermi energy consists of the $p_{x}$ and $p_y$ orbitals of Bi, as its $p_z$ orbital bounds with hydrogens and thus only contributes to bands far away from the Fermi energy. The $p_{x}$ and $p_y$ orbitals on nearby Bi atoms couple via coexisting $\sigma$- and $\pi$- bonds, as shown in Fig.\ref{fig:geomerybandFS}(c) for the nearest-neighbor case. On the experimental side the Bi(111) bilayer has been synthesized recently~\cite{Hirahara2011,Yang2012}. It can be used as a starting template to fabricate the BiH studied here.

Based on the above crystal and band structures, we construct the following $p_{x,y}$-orbital tight-binding model to describe the low energy band structure of the Bi bilayer\cite{Ours1,Ours2,congjun1}.
\begin{eqnarray}
H_0 &=& \sum_{\left\langle i\mu,j\nu\right\rangle\sigma}t_{i\mu,j\nu} c_{i\mu\sigma}^\dag c_{j\nu\sigma} - \lambda  \sum_i c^\dag_i \tau^y s^z c_i. \label{eq:H0}
\end{eqnarray}
 Here $i,j$ label the sites of the honeycomb lattice and $\mu/\nu=x,y$ designate the $p_x$ and $p_y$ orbitals. The hopping integral $t_{i\mu,j\nu}$  can be obtained from the Slater-Koster formula\cite{SlaterKoster}
 \begin{eqnarray}
t_{i\mu,j\nu}=t^{ij}_{\sigma}\cos \theta_{\mu,ij}\cos \theta_{\nu,ij}+
t^{ij}_{\pi}\sin \theta_{\mu,ij}\sin \theta_{\nu,ij},\label{slater_koster}
\end{eqnarray}
where $\theta_{\mu,ij}$ denotes the angle from the $\mu$ direction to $\v r_j-\v r_i$. 
The last term in \Eq{eq:H0} is the only symmetry allowed on-site spin-orbit coupling (SOC), where $\tau$ and $s$ are the orbital and spin Pauli matrices, respectively. In the following, we shall use  $t^{NN}_{\pi}=-0.45t^{NN}_{\sigma}, t^{NNN}_{\sigma}=0, t^{NNN}_{\pi}=-0.15t^{NN}_{\sigma}$ and $\lambda=0.35t^{NN}_{\sigma}$ (for more details see Appendix A). Here the $t^{NN(NNN)}_{\sigma(\pi)}$ denotes the hopping integral via nearest-neighbor (next-nearest-neighbor) $\sigma$- ($\pi$-) bonds respectively.

In Fig.\ref{fig:geomerybandFS}(d) we plot the band structure of \Eq{eq:H0}. A direct band gap $\sim 1.2$ eV opens at the $K$-points due to the large on-site SOC\cite{Ours2}. This gap pushes the conduction band minimum at $K$ above that at $\Gamma$ resulting in an indirect band gap. The above bandstructure captures all main features of the first principle results and is characteristic of all Bi-Hydride/Halide compounds.
It turns out that this bandstructure describes a quantum spin Hall (QSH) insulator.
In Fig.\ref{fig:geomerybandFS}(e) we show the in-gap helical edge modes along the ``zigzag" edges of the buckled honeycomb lattice.

Upon p-type doping, hole-pockets appear around the $K$-points, as shown in Fig.\ref{fig:geomerybandFS}(f) (for 5\% doping). Upon n-type doping, an electron-pocket first appears around $\Gamma$  as shown in Fig.\ref{fig:geomerybandFS}(g) (for 4\% doping). For larger n-type doping additional electron pockets appear around the $K$-points as shown in Fig.~1(h) (for 10\%-doping). 



\subsection{B. The electron correlation.} To describe doped BiH it is important to take the electron correlation into account.
We model the electron correlation by the intra and inter-orbital Coulomb repulsion $U,V$ and the Hund's coupling $J_H$. These parameters are obtained from constrained density functional theory calculations (see Appendix B). 
The total Hamiltonian reads
\bea
H&=&H_0+H_1,\label{total_Hamiltonian}\\
H_1&=&\sum_i \Big\{U (n_{ix\uparrow}n_{ix\downarrow}+n_{iy\uparrow}n_{iy\downarrow})+
V n_{ix}n_{iy}\nonumber\\
&+&J_{H}\big[\sum_{\sigma\sigma^{\prime}}c^\dag_{ix\sigma}c^\dag_{iy\sigma^{\prime}}
c_{ix\sigma^{\prime}}c_{iy\sigma}\nonumber\\&+&\!(c^\dag_{ix\uparrow}c^\dag_{ix\downarrow}
c_{iy\downarrow}c_{iy\uparrow}+h.c.)\big]\Big\}, \label{Hubbard}
\eea
where the usual symmetry argument requires $U=V+2J_{H}$. For BiH, our first principle calculation (Appendix B) yields the following estimates of the interaction parameters: $U=1.16$ eV, $V=0.24$ eV, $J_H=0.46$ eV. 
Upon doping, \Eq{Hubbard} is the starting point of our mean-field Cooper pairing analysis.

\begin{table}
\centering
  \caption{The Cooper pair operators associated with seven different pairing symmetries. Here $c_{a\mu\sigma}$  annihilates an electron with sublattice index
  $a$(=A,B), orbital index $\mu$(=$x,y$) and spin index $\sigma$(=$\uparrow,\downarrow$). The sign $\e_a$ is equal to 1(-1) for sublattice-A(B), respectively.}
\label{Tab:one}
\begin{tabular}{@{}ccccccc@{}}
\\\hline\hline
 symmetry   &  pairing operator   \\
 \hline\hline
 $s$          &  $\sum_{a}\left[c_{ax\uparrow}c_{ay\downarrow}+ c_{ax\downarrow}c_{ay\uparrow}+\beta\sum_{\mu}c_{a\mu\uparrow}c_{a\mu\downarrow}\right]$  \\
 $d_{x^2-y^2}$         &  $\sum_{a}\left(c_{ax\uparrow}c_{ax\downarrow}-c_{ay\uparrow}c_{ay\downarrow}\right)$             \\
 $d_{xy}$         &  $\sum_{a}\left(c_{ax\uparrow}c_{ay\downarrow}-c_{ax\downarrow}c_{ay\uparrow}\right)$          \\
 $f$         &  $\sum_{a}\e_a\left[c_{ax\uparrow}c_{ay\downarrow}+ c_{ax\downarrow}c_{ay\uparrow}+\beta\sum_{\mu}c_{a\mu\uparrow}c_{a\mu\downarrow}\right]$             \\
 $p_x$         &  $\sum_{a}\e_a\left(c_{ax\uparrow}c_{ax\downarrow}-c_{ay\uparrow}c_{ay\downarrow}\right)$             \\
 $p_y$         &  $\sum_{a}\e_a\left(c_{ax\uparrow}c_{ay\downarrow}-c_{ax\downarrow}c_{ay\uparrow}\right)$             \\
 $\left(p\pm ip'\right)_{\uparrow\uparrow,\downarrow\downarrow}$         &  $\sum_{a}\e_ac_{ax\sigma}c_{ay\sigma}$            \\
 \hline\hline
\end{tabular}
\end{table}

\section{III. Mean-field Analysis}
 \subsection{A. Approach: classification and decoupling}
 Due to the absence of Fermi surface(FS) nesting and the relatively weak electron correlation we expect superconducting pairing to be the primary electronic instability. We classify the pairing symmetry according to the transformation property of the gap function under the point group operations.
Because all interaction terms in \Eq{Hubbard} are local, we expect all mean-field superconducting instabilities to be associated with on-site Cooper pairing. There are seven distinct on-site pairing whose Cooper pair operators are given in Table 1. The transformation properties of these operators under the point group operations are given in the Appendix C.

In terms of local pairing operators Eq.(\ref{Hubbard}) can be expressed as the sum of an interorbital pairing part
\begin{eqnarray}
H_{\rm inter}&=&\left(V+J_H\right)\sum_{i}\hat{\Delta}^
{\left(-\right)\dagger}_{x y\uparrow\downarrow}\left(i\right)\hat{\Delta}^
{\left(-\right)}_{x y\uparrow\downarrow}\left(i\right)\nonumber\\&+&\left(V-J_H\right)
\Big[\sum_{i\sigma}\hat{\Delta}^
{\dagger}_{x y\sigma\sigma}\left(i\right)\hat{\Delta}_{x y\sigma\sigma}\left(i\right)\nonumber \\&+&
\sum_{i}\hat{\Delta}^
{\left(+\right)\dagger}_{x y\uparrow\downarrow}\left(i\right)\hat{\Delta}^
{\left(+\right)}_{x y\uparrow\downarrow}\left(i\right)\Big] \label{decouple1}
\end{eqnarray}
and an intraorbital pairing part
\begin{eqnarray}
H_{\rm intra}&=&U\sum_{i\mu}\hat{\Delta}^{\dagger}_{\mu\mu\uparrow\downarrow}\left(i\right)\hat{\Delta}_{\mu\mu\uparrow\downarrow}\left(i\right)
\nonumber\\&+&J_H\sum_{i,\mu\ne\nu}\hat{\Delta}^{\dagger}_{\mu\mu\uparrow\downarrow}\left(i\right) \hat{\Delta}_{\nu\nu\uparrow\downarrow}\left(i\right),\label{decouple2}
\end{eqnarray}
where $\hat{\Delta}_{\mu\nu\sigma\sigma^{\prime}}(i)\equiv c_{i\mu\sigma}c_{i\nu\sigma^{\prime}}$ and $\hat{\Delta}^{\left(\pm\right)}_{\mu\nu\uparrow\downarrow}\equiv \frac{1}{\sqrt{2}}\left(\hat{\Delta}_{\mu\nu\uparrow\downarrow}\pm\hat{\Delta}_{\mu\nu\downarrow\uparrow}\right)$.
From (\ref{decouple1}) and (\ref{decouple2}) it's clear that for $U,V>0$ and $J_H>V$, the energetically favored pairing channels include $\hat{\Delta}_{x y\sigma\sigma}$ and $\hat{\Delta}^{\left(+\right)}_{x y\uparrow\downarrow}$.  These channels include pairings with  $s$, $f$ and $p+ip'$ symmetry in Table 1.  (Note that although the intraorbital pairing is energetically unfavored, the SOC can mix them with the inter-orbital pairing as long as symmetry allows it.)
\begin{figure*}
\includegraphics[width=7in]{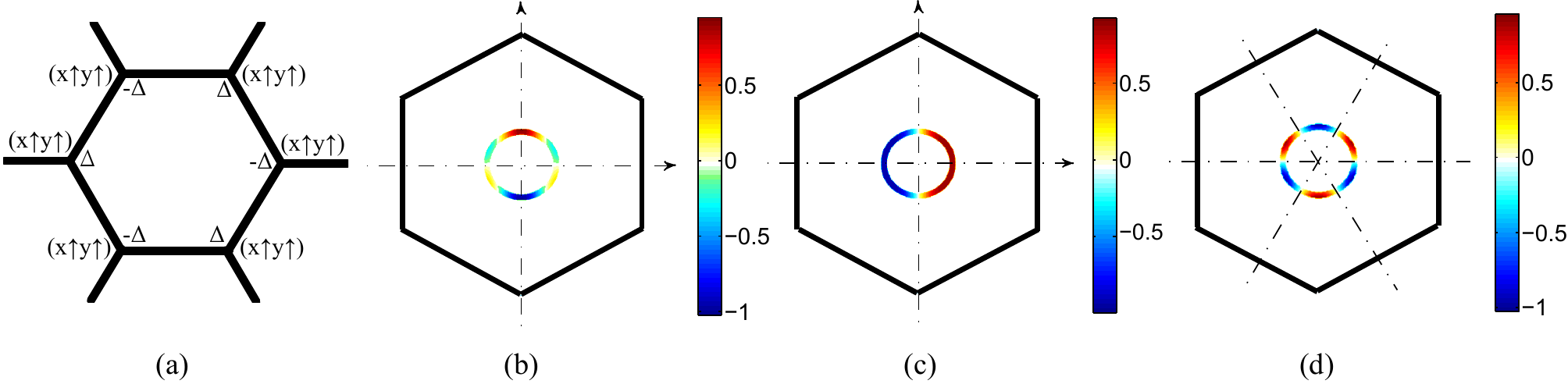}
\caption{(color online). (a) The $(p+ip')_{\uparrow\uparrow}$ gap function in real-space. The real (b) and imaginary (c) parts of the $(p+ip')_{\uparrow\uparrow}$ gap function and the f-wave gap function (d), for  4\% electron doping, plotted around the FS.}\label{fig:gap}
\end{figure*}

Through mean-field decoupling of Eq.(\ref{decouple1}) and Eq.(\ref{decouple2}) in the seven possible pairing channels listed in Table 1, we obtain the mean-field Hamiltonian for different pairing symmetries. Solving the ground state of the mean-field Hamiltonian self-consistently, we derive the gap equation for each pairing channel. Solving the gap equations and compare the associated mean-field ground state energies, we determine the leading pairing symmetry. In our calculation, we also perform the mean-field decoupling in the momentum space, which yields the same results but is more convenient in practise. Upon Fourier transform, the Hamiltonian reads
\begin{eqnarray}
H&=&\sum_{\mathbf{k}\s\mu\nu}h_{\mu\nu\s}\left(\mathbf{k}\right)
c^{\dagger}_{\mathbf{k}\mu\sigma}c_{\mathbf{k}\nu\sigma}\nonumber\\&+&\sum_{i\s_1\s_2\mu\nu\theta\xi}U
^{\mu\nu\sigma_{1}}_{\theta\xi\sigma_{2}}c^{\dagger}_{i\mu\sigma_{1}}
c_{i\nu\sigma_{1}}c^{\dagger}_{i\theta\sigma_{2}}c_{i\xi\sigma_{2}}\nonumber\\
&\to&\sum_{\mathbf{k}\alpha\s}\varepsilon_{\mathbf{k}\s}^{\alpha}c^{\dagger}_{\mathbf{k}\alpha\sigma}c_{\mathbf{k}\alpha\sigma}\nonumber\\&+&\sum_{
\mathbf{k}\mathbf{q}\s_1\s_2\mu\nu\theta\xi}U
^{\mu\nu\sigma_{1}}_{\theta\xi\sigma_{2}}c^{\dagger}_{\mathbf{k}\mu\sigma_{1}}
c_{\mathbf{q}\nu\sigma_{1}}c^{\dagger}_{\mathbf{-k}\theta\sigma_{2}}c_{\mathbf{-q}\xi\sigma_{2}}.\label{k_space_model}
\end{eqnarray}
Here $\mu/\nu=1,\cdots,4$ represent for orbital-sublattice indices (with 1(2) and 3(4) representing for the $p_x$($p_y$) orbitals on sublattice-A and B respectively), and $\alpha=1,\cdots,4$ are the band indices. The transformation between the orbital basis
$c_{\mathbf{k}\mu\sigma} $ and the band eigen basis $c_{\mathbf{k}\alpha\sigma}$ is
given by,
\begin{equation}
c_{\mathbf{k}\mu\sigma}=\sum_{\alpha}\xi_{\mu\alpha}^{\s}\left({\mathbf{k}}\right)c_{\mathbf{k}\alpha\sigma}.
\end{equation}
Note that due to the time-reversal symmetry, we have the following relations,
\begin{eqnarray}
\varepsilon_{\mathbf{k}\uparrow}^{\alpha}&=&\varepsilon_{\mathbf{-k}\downarrow}^{\alpha}\nonumber\\
\xi^{\alpha}\left({\mathbf{k}\uparrow}\right)&=&\xi^{\alpha,\star}\left({\mathbf{-k}\downarrow}\right),\label{sym}
\end{eqnarray}

The interacting parameter $U ^{\mu\nu\sigma_{1}}_{\theta\xi\sigma_{2}}$ is defined as,
\begin{eqnarray}
U^{l_{1}l_{2}\s}_{l_{3}l_{4}\s}&=&\left\{\begin{array}{cc}{\frac{V-J_{H}}{4},l_{1}=l_{2}\ne
l_{3}=l_{4}\in\{1,2\}or\{3,4\}}\\{\frac{J_{H}-V}{4},l_{1}=l_{4}\ne
l_{3}=l_{2}\in\{1,2\}or\{3,4\}}\end{array}\right.\nonumber\\U^{l_{1}l_{2}\s}_{l_{3}l_{4}\bar{\s}}&=&\left\{\begin{array}{cc}{\frac{U}{2},l_{1}=l_{2}=l_{3}=l_{4}\in\{1,2\}or\{3,4\}}\\{\frac{V}{2},l_{1}=l_{2}\ne
l_{3}=l_{4}\in\{1,2\}or\{3,4\}}\\{\frac{J_{H}}{2},l_{1}=l_{3}\ne
l_{2}=l_{4}\in\{1,2\}or\{3,4\}}\\{\frac{J_{H}}{2},l_{1}=l_{4}\ne
l_{3}=l_{2}\in\{1,2\}or\{3,4\}}\end{array}\right.
\end{eqnarray}
Due to the local nature of the Hubbard interaction, in order to get a non-zero interaction parameter, the indices $l_{i} (i=1,\cdots,4)$ have to either all belong to the set $\{1,2\}$ or all belong to the set $\{3,4\}$. Moreover the fermion anticommutation relation implies,
\begin{eqnarray}
U^{\mu\nu\s}_{\theta\xi\s}&=&U^{\theta\xi\s}_{\mu\nu\s}=-U^{\mu\xi\s}_{\theta\nu\s}=-U^{\theta\nu\s}_{\mu\xi\s},\nonumber\\
U^{\mu\nu\s}_{\theta\xi\bar{\s}}&=&U^{\theta\xi\s}_{\mu\nu\bar{\s}}.\label{symmetry_2}
\end{eqnarray}
Note that in the above we have restricted ourselves to the Cooper scattering channel.

In the weak pairing limit, only intra-band pairing needs to be considered. In that case
\begin{eqnarray}
H_{\rm{I}}&=&\sum_{\mathbf{k}\mathbf{q}\mu\nu\theta\xi\s_1\s_2}U
^{\mu\nu\sigma_{1}}_{\theta\xi\sigma_{2}}c^{\dagger}_{\mathbf{k}\mu\sigma_{1}}
c_{\mathbf{q}\nu\sigma_{1}}c^{\dagger}_{\mathbf{-k}\theta\sigma_{2}}c_{\mathbf{-q}\xi\sigma_{2}}\nonumber\\
&\to&\sum_{\mathbf{k}\mathbf{q}\alpha\beta\s_1\s_2}V
_{\alpha\beta}^{\sigma_{1}\s_2}\left(\mathbf{k},\mathbf{q}\right)c^{\dagger}_{\mathbf{k}\alpha\sigma_{1}}
c_{\mathbf{q}\beta\sigma_{1}}c^{\dagger}_{\mathbf{-k}\alpha\sigma_{2}}c_{\mathbf{-q}\beta\sigma_{2}},\nonumber\\
\end{eqnarray}
where parameter
\begin{eqnarray}
V_{\alpha\beta}^{\sigma_{1}\s_2}\left(\mathbf{k},\mathbf{q}\right)&=&\sum_{\mu\nu\theta\xi}U
^{\mu\nu\sigma_{1}}_{\theta\xi\sigma_{2}}\xi^{\s_1,*}_{\mu\alpha}\left(\mathbf{k}\right)
\xi^{\s_1}_{\nu\beta}\left(\mathbf{q}\right)\nonumber\\&\times&\xi^{\s_2,*}_{\theta\alpha}\left(\mathbf{-k}\right)\xi^{\s_2}_{\xi\beta}\left(\mathbf{-q}\right).\label{v}
\end{eqnarray}
Due to the $S_z$-conservation and the inversion symmetry, the pairing potential $V_{\alpha\beta}^{(i)}\left(\mathbf{k},\mathbf{q}\right)$ can take the following four possible forms,
\begin{eqnarray}
V^{(1,1)}_{\alpha\beta}(\mathbf{k,q})&\equiv&  V^{\uparrow\uparrow}_{\alpha\beta}(\mathbf{k,q})\nonumber\\
V^{(1,-1)}_{\alpha\beta}(\mathbf{k,q})&\equiv&  V^{\downarrow\downarrow}_{\alpha\beta}(\mathbf{k,q})\nonumber\\
  V^{(1,0)}_{\alpha\beta}(\mathbf{k,q})&\equiv&  \frac{1}{2}\left[V^{\uparrow\downarrow}_{\alpha\beta}(\mathbf{k,q})-V^{\uparrow\downarrow}_{\alpha\beta}(\mathbf{k,-q})\right]\nonumber\\
 V^{(0,0)}_{\alpha\beta}(\mathbf{k,q})&\equiv& \frac{1}{2}\left[V^{\uparrow\downarrow}_{\alpha\beta}(\mathbf{k,q})+V^{\uparrow\downarrow}_{\alpha\beta}(\mathbf{k,-q})\right]\end{eqnarray}
The first three channels are for odd-parity pairings with $S_z=1,-1,0$ respectively. The last channel is for even-parity. Particularly, from equation (\ref{sym}) and (\ref{v}), we found
\begin{equation}
V^{\uparrow\uparrow}_{\alpha\beta}(\mathbf{k,q})=V^{\downarrow\downarrow,*}_{\alpha\beta}(\mathbf{-k,-q}).\label{conjg}
\end{equation}

To determine $T_c$, we use the following linearized gap equation for each pairing channel,
\begin{equation}
 -\frac{1}{(2\pi)^2}\sum_{\beta}\oint_{FS}
dk'_{\Vert}\frac{V^{(i)}_{\alpha\beta}(\mathbf{k,k'})}{v^{\beta}_{F}(\mathbf{k'})}\Delta_{\beta}(\mathbf{k'})=r
 \Delta_{\alpha}(\mathbf{k}).\label{eigenvalue_Tc}
\end{equation}
Here, $\beta$ labels the FS and the integral is performed around each connected FS. Moreover, $v^{\beta}_{F}(\mathbf{k'})$
is the Fermi velocity at $\mathbf{k'}$ on the $\beta-$th FS, and
$k'_{\Vert}$ represents the tangential component of $\mathbf{k}'$. Solving Eq.(\ref{eigenvalue_Tc}) as an
eigenvalue problem, we obtain the pairing eigenvalue $r$ ($r$ is related to the superconducting critical temperature $T_c$ via $T_{c}\sim {\rm cutoff~energy}$ $e^{-1/r}$) and gap function $\Delta_{\alpha}(\mathbf{k})$. The leading gap function is the one corresponds to
the largest eigenvalue $r$.

\subsection{B. Results}
Since our main interest is TRI topological superconductivity and in the weak pairing limit this type of superconductivity requires the FS to enclose the TRI momenta, in the following we shall focus only on low n-type doping ($<8\%$) regime, where the fermiology resembles that of Fig.\ref{fig:geomerybandFS}(g).





\begin{figure*}
\includegraphics[width=5in]{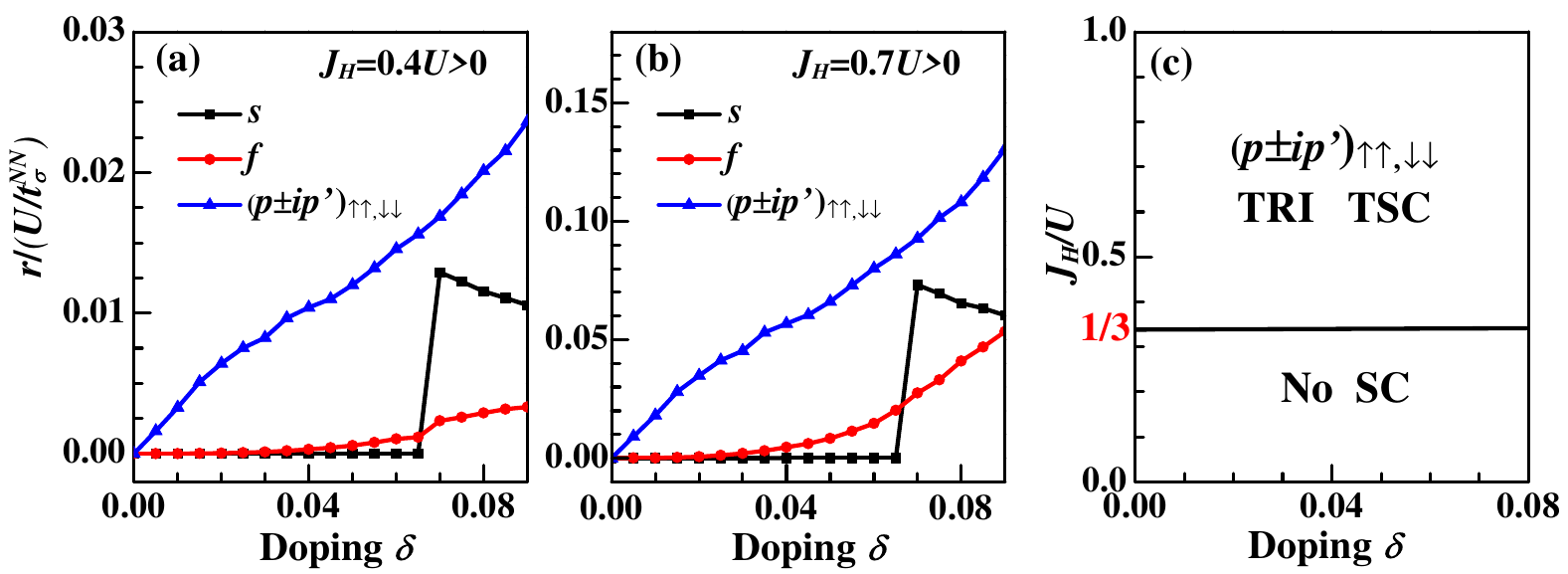}
\caption{(color online). Doping-dependence of the pairing eigenvalue $r$ over $U/t^{NN}_{\sigma}$ for several leading pairing symmetries. (a) $J_{H}=0.4 U>0$, (b) $J_H=0.7U>0$. (c). The phase-diagram for positive $U$. 
}\label{fig:dop}
\end{figure*}

The result of our mean-field analysis shows  that for $J_H>\frac{U}{3}$ (which is equivalent to $J_H>V$ upon using the relation $V=U-2 J_H$) and for doping concentration $<8\%$, the leading pairing instability is the interorbital equal spin, $(p+ip')_{\uparrow\uparrow}$ and  $(p-ip')_{\downarrow\downarrow}$, pairing (henceforth abbreviated as $(p\pm ip')_{\uparrow\uparrow,\downarrow\downarrow}$). This is listed in the 7th row of Table 1. The real space gap function for the  $\uparrow\uparrow$ pairing is shown in Fig.\ref{fig:gap}(a). The real and imaginary part of the same gap function plotted around the FS (for doping $\delta=0.04$) are shown in Fig.\ref{fig:gap}(b) and Fig.\ref{fig:gap}(c), respectively. From Fig.\ref{fig:gap}(b,c) it is apparent that the real (imaginary) part of the gap function has the $p_y$ ($p_x$) symmetry. The momentum space $\downarrow\downarrow$ gap function is the complex conjugate of that for $\uparrow\uparrow$ so that TR symmetry is respected.  


The doping-dependence of the eigenvalue $r$ of the linearized gap equation Eq.(\ref{eigenvalue_Tc}) for various pairing symmetries is shown in Fig.\ref{fig:dop}(a). The interaction parameter used to construct this figure is $J_H=0.4U$ ($V=U-2 J_H$) appropriate for BiH. From this figure it is apparent that $(p\pm ip')_{\uparrow\uparrow,\downarrow\downarrow}$ is the leading pairing symmetry for doping $< 8\%$. In the phase-diagram shown in Fig.\ref{fig:dop}(c), we determine the leading pairing symmetry as a function of $J_H/U$ and doping level $\delta$. The result shows $(p\pm ip')_{\uparrow\uparrow,\downarrow\downarrow}$ pairing is realized for low doping and  $U>J_H>U/3$. We emphasize that SOC plays a crucial role in stabilizing the $(p\pm ip')_{\uparrow\uparrow,\downarrow\downarrow}$ pairing. When the SOC parameter $\lambda$ is set to zero, the leading pairing symmetry becomes $f$-wave (shown in Fig.\ref{fig:gap}(d)).



In the $(p\pm ip')_{\uparrow\uparrow,\downarrow\downarrow}$ superconducting state there are gapless helical Majorana edge modes (marked by ``A'' in Fig.\ref{fig:edge}(a) for the ``zigzag" edge of the 4\% n-type doped system).  
The gapless modes centered at large momenta, marked by ``B", are the remnant of the complex fermion helical edge modes of the QSH insulator. Such edge modes can not participate in the $(p\pm ip')_{\uparrow\uparrow,\downarrow\downarrow}$ pairing because the partners of a Cooper pair are localized on opposite edges. Since  the pairing interactions in \Eq{decouple1} and \Eq{decouple2} are completely local, such distant pairing can not occur. Thus we have an interesting situation where the edge modes of the $(p\pm ip')_{\uparrow\uparrow,\downarrow\downarrow}$ superconductor are composed of the complex fermion edge modes of the parent QSH insulator and the superconducting helical Majorana edge modes. By degree of freedom counting these amount to five Majorana fermion modes per edge which can not be gapped out by TRI perturbations.\\

\section{ Discussion and conclusion}

 \noindent In addition to the electron-electron interaction, phonon can play an substantial role in Cooper pairing, especially for weakly correlated materials. In general phonon mediates additional attractive interaction which can reduce the value of $U$ and $V$ in the preceding  discussions. When such extra attraction  is weak so that $U$ is still positive ($V$ can have either sign), the main effect is to increase the $J_H/U$ ratio. In Fig.\ref{fig:dop}(b), the doping dependence of the pairing eigenvalue $r$ for the three leading pairing symmetries is shown for, e.g., $J_H=0.7U$ and $V=-0.4U$. The result shows $(p\pm ip')_{\uparrow\uparrow,\downarrow\downarrow}$ remains the leading pairing symmetry. The only difference is the $T_c$ enhancement. Thus $(p\pm ip')_{\uparrow\uparrow,\downarrow\downarrow}$ can survive weak phonon-mediated attractive interaction.

\begin{figure}
\includegraphics[scale=0.42]{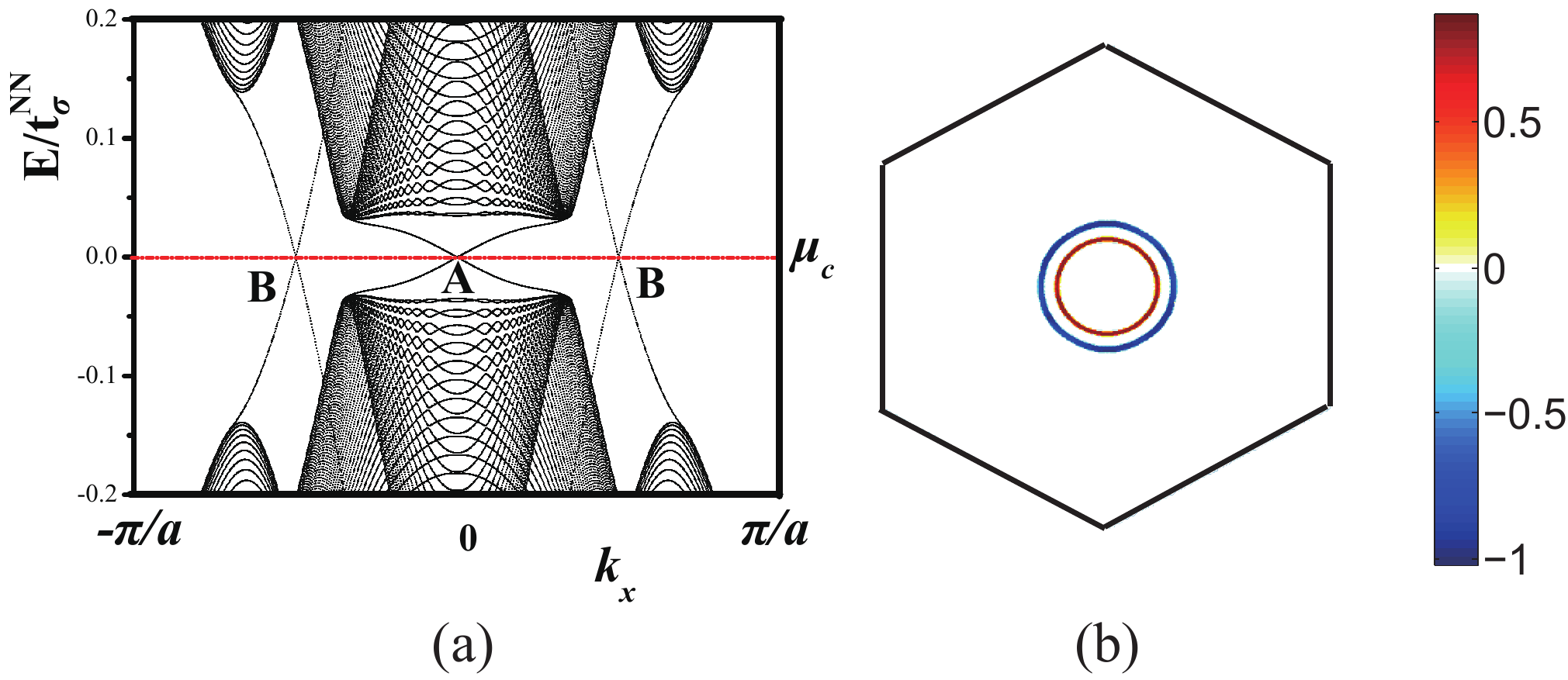}
\caption{(color online).(a) The edge spectrum (for the ``zigzag" edges) for the $(p\pm ip')_{\uparrow\uparrow,\downarrow\downarrow}$ superconducting state at 4\%-doping. The in-gap states marked by ``A" are helical Majorana modes. The in-gap states marked by ``B" are helical complex fermion modes. They are the remnant of the edge modes of the parent QSH insulator. (b) The leading gap function on the Rashba-split Fermi surfaces for 4\% electron doping. (The Rashba interaction used to construct this figure is $\lambda_{R}=0.08t^{NN}_{\sigma}$).} \label{fig:edge}
\end{figure}

 Finally, we discuss the effect of the inevitable inversion symmetry breaking caused by doping. Such breaking  of symmetry will induce the Rashba SOC in the Hamiltonian. The Rashba interaction splits the spin degenerate FS into spin non-degenerate ones.  In Fig.\ref{fig:edge}(b) we show the 4\% n-type doped gap function after the FS is split by a weak Rashba SOC $\lambda_{R}=0.08t^{NN}_{\sigma}$.  The opposite sign of the gap function on the two FS implies the pairing remains TRI and topological\cite{Xiaoliang_criterion}. The fact that a centro-symmetric TRI topological SC can evolve smoothly into a non centro-symmetric TRI topological SC upon the introduction of weak Rashba coupling has been discussed in Ref.\cite{Qianghua}. As a function of doping the TRI topological pairing remains until $\delta>0.08$ where the Rashba SOC causes gap nodes on the split FS.


In conclusion by combining first principle and mean-field calculations, we predict a single bilayer of hydrogenated Bismuth, BiH, to be a TRI topological superconductor. Moreover we predict the edge modes of such superconductor consist of a pair of helical Majorana fermion and two pairs of helical complex fermion edge modes. We believe the physics discussed here can also occur in other bilayer Bi-Hydride/Halide compounds.

 {\it Acknowledgements:} We are grateful to Ming-Cui Ding to provide us the interaction parameters of BiH. We thank Hong Yao, Fan Zhang, Zi-Yang Meng, Yi-Fan Jiang and Li-Da Zhang for helpful discussions and to Yong-You Zhang for the help of figure preparation. Yugui Yao is supported by the MOST Project of China (Grants Nos. 2014CB920903, 2011CBA00100) , the NSFC (Grant Nos.11174337, and 11225418) and Specialized Research Fund for the Doctoral Program of Higher Education of China (Grants No. 20121101110046). Fan Yang is supported in part by NSFC under the grant NO. 11274041 and No.11334012 and by the NCET program under the grant No. NCET-12-0038. Cheng-Cheng Liu is supported by NSFC under the grant NO. 11404022. Yu-Zhong Zhang is supported by NSFC under the grant NO.11174219.  Dung-Hai Lee is supported by DOE Office of Basic Energy Sciences, Division of Materials Science, grant DE-AC02-05CH11231.

\begin{widetext}

\appendix

\section{Appendix A: The tight-binding parameters}

A low energy four band tight binding Hamiltonian with the on-site SOC interaction is given by Eq.(1) of the main text. The Slater-Koster tight-binding (TB) parameters~\cite{SlaterKoster} are obtained by least square fitting the tight binding bandstructure to that obtained from the first principle calculation\cite{Kresse1996}. By including up to the next-nearest-neighbor (NNN) hoppings, our tight-binding model reproduces the first principle energy bands fairly well. This is shown in Fig.\ref{fig:geomerybandFS}(d) of the main text. Similarly, we obtain the Slater-Koster parameters for the Bi-Halide bilayer.
The tight-binding parameters in units of $t^{NN}_{\sigma}$ for the bilayer BiH and other Bi Halide are listed in Table~\ref{tab:hopping-parameter}.

\section{Appendix B: The estimates of the Interaction parameters}

\subsection{a. The interaction parameter J$_H$}\label{HundJ}

Because the solid state screening of the Slater integrals F$^2$ and F$^4$ is usually small, we estimate the Hund's coupling J$_H$ by taking the atomic limit. For this purpose, we construct four simple cubic lattices composed purely of the Bi atoms with large lattice constants of $10$, $20$, $40$, and $80 \AA$. The full potential nonorthogonal local orbital (FPLO) code~\cite{Koepernik,Opahle} within both local density approximation (LDA) and generalized gradient approximation(GGA) are employed to calculate the difference of the total energies between two magnetic states with different total Bi moment ( $1\mu_B$ and $3\mu_B$) for each artificial crystal discussed above. When the total moment is $3 \mu_B$, the spins of the three electrons in the 6p orbitals of each Bi atom are aligned. In contrast, when the total moment is $1 \mu_B$, one of the three electrons has spin opposite to the other two. Assuming the above energy difference is due to the Hund's rule interaction, namely, $J_H=(E(m=1)-E(m=3))/2$, we obtain the results shown in Table~\ref{Tab:one}. From Table~\ref{Tab:one}, it is clear that the obtained $J_H$ has little dependence on the lattice parameters of the artificial crystal.
\begin{table}
\centering
\caption{The parameters $t^{NN}_{\sigma}$ (in unit of eV) as well as the ratios $t^{NN}_{\pi}/t^{NN}_{\sigma}$, $t^{NNN}_{\sigma}/t^{NN}_{\sigma}$, $t^{NNN}_{\pi}/t^{NN}_{\sigma}$, and $\lambda/t^{NN}_{\sigma}$ for the 2D Bi-Hydride/Halide bilayer family, which are obtained by fitting with the FP calculations. Note that $\lambda=E_{g}/2$, with $E_{g}$ to be the gap opened by SOC at the Dirac points.}\label{tab:hopping-parameter}
\begin{tabular}{cccccccccccc}
\\\hline\hline
& system & $t^{NN}_{\sigma}$(eV) & $t^{NN}_{\pi}/t^{NN}_{\sigma}$ & $t^{NNN}_{\sigma}/t^{NN}_{\sigma}$ & $t^{NNN}_{\pi}/t^{NN}_{\sigma}$ & $\lambda/t^{NN}_{\sigma}$  \\
 \hline\hline
& BiH  & 1.79 & -0.45 &  0.04 & -0.15 &  0.35\\
& BiF  & 1.51 & -0.36 &  0.15 & -0.13 &  0.36\\
& BiCl & 1.43 & -0.39 &  0.10 & -0.13 &  0.39\\
& BiBr & 1.39 & -0.40 &  0.09 & -0.11 &  0.48\\
& BiI  & 1.31 & -0.44 &  0.08 & -0.10 &  0.50\\
 \hline\hline
\end{tabular}
\end{table}
\subsection{b. The interaction parameter U}\label{CoulU}

Constrained LDA and GGA calculations \cite{Anisomov,Madsen}implemented in the Wien2k code are carried out to estimate the
screened Coulomb intergral $F^0_{eff}=U-J_H$\cite{Anisomov,Madsen}, which corresponds to the difference between the electron affinity and the ionization energy upon adding and removing an electron in the valence shell of a given atom.
Such constrained LDA and GGA approaches are based on the observation that the energy of a system with increased or reduced particle number is in principle computable within density functional theory. $F^0_{eff}$ is then defined as the derivative of the total energy with respect to the constrained occupation number on a given shell. Moreover, in order to avoid double counting of the hopping matrix elements in the calculation of $F^0_{eff}$,
one has to construct a supercell and set one of the atom as an impurity where hopping to the rest of the system is suppressed.

In our calculations, we constructed a superlattice which consists of $2 \times 2 \times 1$ supercell where 8 Bi ions and 8 H ion are included. One of the Bi ion is set as the impurity where two 6p electrons are constrained to the core state to avoid the hopping to the rest of the system.

The results are shown in Table~\ref{Tab:two}.

\begin{table}
\centering
  \caption{The Hund's coupling J$_H$ of 6p orbital of Bi atom in a simple cubic lattice with large lattice constant. J$_H$ is estimated from the energy difference between two magnetic states with magnetic moments of $m=1$ and $m=3$. The groundstate energies of the magnetically ordered states are obtained from first principles calculations by FPLO code. Both LDA and GGA are used for comparison.}
\label{Tab:one}
\begin{tabular}{@{}ccccccc@{}}
\\\hline\hline
 a($A^{o}$)   &  J$^{LDA}_H$(eV) & J$^{GGA}_H$(eV)  \\\hline\hline
 80         &  0.455           & 0.563  \\
 40         &  0.455           & 0.562  \\
 20         &  0.455           & 0.562  \\
 10         &  0.456           & 0.562  \\
 \hline\hline
\end{tabular}
\end{table}
\begin{table}
  \caption{The results of $U$ from constraint LDA and GGA calculations.}
\label{Tab:two}
\begin{tabular}{@{}ccccccc@{}}
\\\hline\hline
&  LDA & GGA  \\\hline\hline
$U=F^0_{eff}+J_H (eV)$        &  1.16& 1.49  \\
\hline\hline
\end{tabular}
\end{table}

\section{Appendix C: Pairing symmetry}
The point-group symmetry of the system includes the following operations: the inversion $\hat{I}$, the reflection $\hat{R}_{x,y}$ about the $x$- or $y$-axes (here the $x,y$-axes are shown in Fig.\ref{fig:geomerybandFS}(c) of the main text), and the rotation $\hat{R}(\theta)$ about the $z$-axis with the angle $\theta=\pm\frac{\pi}{3},\pm\frac{2\pi}{3}$. Note that in the real material, the point-group is $D_{3d}$ instead of $D_{6h}$, because the material is buckled and the A and B sublattices locate in different planes. In this point-group, an extra $xy$ plane mirror reflection should be added to the $\hat{R}_y$ reflection and the $\pm\frac{\pi}{3}$ rotations since these operations change the sublattice index of a lattice site. Thus in the definition of the point-group operations below, we have added the extra $xy$ plane mirror reflection where it is needed. Under these operations, an electron operator $c_{a\mu\sigma}$ (with $a,\mu,\sigma$ labeling the sublattice, orbital and spin degrees of freedom ) transforms as,
\begin{eqnarray}
\hat{I}&:&c_{a\mu\sigma}\to -c_{\bar{a}\mu\sigma} \nonumber\\
\hat{R}_x&:&c_{a\mu\sigma}\to (-1)^{\mu}\sigma c_{a\mu\bar{\sigma}}\nonumber\\
\hat{R}_y&:&c_{a\mu\sigma}\to (-1)^{\mu+1}\sigma c_{\bar{a}\mu\bar{\sigma}}\nonumber\\
\hat{R}(\theta)&:&c_{a x\sigma}\to e^{-i\sigma\theta}\left(c_{a' x\sigma}\cos\theta+c_{a' y\sigma}\sin\theta\right)\nonumber\\
\hat{R}(\theta)&:&c_{a y\sigma}\to e^{-i\sigma\theta}\left(-c_{a' x\sigma}\sin\theta+c_{a' y\sigma}\cos\theta\right),\label{operation}
\end{eqnarray}
where
\begin{eqnarray}
(-1)^{\mu}&=&\left\{\begin{array}{cc}{1,\mu=x} \\{-1,\mu=y}\end{array}\right.\nonumber\\
\sigma&=&\left\{\begin{array}{cc}{1,\sigma=\uparrow} \\{-1,\sigma=\downarrow}\end{array}\right.\nonumber\\
a'&=&\left\{\begin{array}{cc}{\bar{a},\theta=\pm\pi/3} \\{a,\theta=\pm2\pi/3}\end{array}\right.\label{pm}
\end{eqnarray}
Here, $\bar{a}$ designates the opposite sublattice of $a$. Before going through the detailed transformation property of each pair operator, we note the following two points

(1).To determine the inversion parity, one can simply examine whether the pair operator on the two sublattices take the same form or differ by a sign. Therefore we shall focus on the other symmetry operations in the following.

(2). The factor $e^{-i\sigma\theta}$ is equal to unity for the total $S_z=0$ Cooper pairs.

Now we check all seven pairing symmetries one by one.

\subsection{a. s-wave}
There can be two types of s-wave: one is inter-orbital and the other is intra-orbital, we analyze them separately.
\subsubsection{1. inter-orbital s-wave}
The pair operator associated with the inter-orbital s-wave is given by
\begin{equation}
\hat{\Delta}^{inter}_{s}\equiv \sum_{a}c_{ax\uparrow}c_{ay\downarrow}+ c_{ax\downarrow}c_{ay\uparrow}.\label{inter_s}
\end{equation}
Acted by $\hat{R}_{x,y}$ and $\hat{R}(\theta)$, it transforms as
\begin{eqnarray}
\hat{R}_{x}&:&\hat{\Delta}^{inter}_{s}\to (-1)^{0+1}(-1)^{0+1}\left(\sum_{a}c_{ax\downarrow}c_{ay\uparrow}+ c_{ax\uparrow}c_{ay\downarrow}\right)=\hat{\Delta}^{inter}_{s}\nonumber\\
\hat{R}_{y}&:&\hat{\Delta}^{inter}_{s}\to (-1)^{1+0}(-1)^{0+1}\left(\sum_{a}c_{\bar{a}x\downarrow}c_{\bar{a}y\uparrow}+ c_{\bar{a}x\uparrow}c_{\bar{a}y\downarrow}\right)=\hat{\Delta}^{inter}_{s}\nonumber\\
\hat{R}(\theta)&:&\hat{\Delta}^{inter}_{s}\to \sum_{a}\left(\cos \theta c_{a'x\uparrow}+\sin \theta c_{a'y\uparrow}\right)\left(-\sin \theta c_{a'x\downarrow}+\cos \theta c_{a'y\downarrow}\right)\nonumber\\&&+\left(\cos \theta c_{a'x\downarrow}+\sin \theta c_{a'y\downarrow}\right)\left(-\sin \theta c_{a'x\uparrow}+\cos \theta c_{a'y\uparrow}\right)\nonumber\\&&=\sum_{a}c_{a'x\uparrow}c_{a'y\downarrow}+ c_{a'x\downarrow}c_{a'y\uparrow}=\hat{\Delta}^{inter}_{s}
\end{eqnarray}
In addition, this pair operator is obviously inversion even since on the A and B sublattices it has the same form. Thus it has s-wave symmetry.

\subsubsection{2. intra-orbital s-wave}
The pair operator associated with the intra-orbital s-wave is given by,
\begin{equation}
\hat{\Delta}^{intra}_{s}\equiv \sum_{a\mu}c_{a\mu\uparrow}c_{a\mu\downarrow}.\label{intra_s}
\end{equation}
When acted by $\hat{R}_{x,y}$ and $\hat{R}(\theta)$, it transforms as
\begin{eqnarray}
\hat{R}_{x}&:&\hat{\Delta}^{intra}_{s}\to \sum_{a\mu}(-1)^{\mu+\mu}(-1)^{0+1}c_{a\mu\downarrow}c_{a\mu\uparrow}=\sum_{a\mu}c_{a\mu\uparrow}c_{a\mu\downarrow}=\hat{\Delta}^{intra}_{s}\nonumber\\
\hat{R}_{y}&:&\hat{\Delta}^{intra}_{s}\to \sum_{a\mu}(-1)^{\mu+\mu+2}(-1)^{0+1}c_{\bar{a}\mu\downarrow}c_{\bar{a}\mu\uparrow}=\sum_{a\mu}c_{\bar{a}\mu\uparrow}c_{\bar{a}\mu\downarrow}=\hat{\Delta}^{intra}_{s}\nonumber\\
\hat{R}(\theta)&:&\hat{\Delta}^{intra}_{s}\to \sum_{a}\left(\cos \theta c_{a'x\uparrow}+\sin \theta c_{a'y\uparrow}\right)\left(\cos \theta c_{a'x\downarrow}+\sin \theta c_{a'y\downarrow}\right)\nonumber\\&&+\left(-\sin \theta c_{a'x\uparrow}+\cos \theta c_{a'y\uparrow}\right)\left(-\sin \theta c_{a'x\downarrow}+\cos \theta c_{a'y\downarrow}\right)\nonumber\\&&=\sum_{a}c_{a'x\uparrow}c_{a'x\downarrow}+ c_{a'y\uparrow}c_{a'y\downarrow}=\hat{\Delta}^{intra}_{s}
\end{eqnarray}
In addition, this pair operator is obviously inversion even. Thus this pair operator has s-wave symmetry.

Since the inter-orbital and intra-orbital s-wave pair operators transform identically, they are allowed to mix as
\begin{equation}
\hat{\Delta}_{s}\equiv \sum_{a}\left(c_{ax\uparrow}c_{ay\downarrow}+ c_{ax\downarrow}c_{ay\uparrow}\right)+\beta\sum_{a\mu}c_{a\mu\uparrow}c_{a\mu\downarrow}
\end{equation}

\subsection{b. d-wave}
There are two types of d-wave pairings, i.e. $\Delta_{d_{x^2-y^2}}$ and $\Delta_{d_{xy}}$, which form a 2D representation of the point-group. In the following, we check the transformation properties of them separately.
\subsubsection{1. $d_{xy}$-wave}
The pair operator is given by
\begin{equation}
\hat{\Delta}_{d_{xy}}\equiv \sum_{a}c_{ax\uparrow}c_{ay\downarrow}-c_{ax\downarrow}c_{ay\uparrow}.\label{inter_s}
\end{equation}
When acted upon by $\hat{R}_{x,y}$ and $\hat{R}(\theta)$, it transforms as
\begin{eqnarray}
\hat{R}_{x}&:&\hat{\Delta}_{d_{xy}}\to (-1)^{0+1}(-1)^{0+1}\left(\sum_{a}c_{ax\downarrow}c_{ay\uparrow}- c_{ax\uparrow}c_{ay\downarrow}\right)=-\hat{\Delta}_{d_{xy}}\nonumber\\
\hat{R}_{y}&:&\hat{\Delta}_{d_{xy}}\to (-1)^{1+0}(-1)^{0+1}\left(\sum_{a}c_{\bar{a}x\downarrow}c_{\bar{a}y\uparrow}- c_{\bar{a}x\uparrow}c_{\bar{a}y\downarrow}\right)=-\hat{\Delta}_{d_{xy}}\nonumber\\
\hat{R}(\theta)&:&\hat{\Delta}_{d_{xy}}\to \sum_{a}\left(\cos \theta c_{a'x\uparrow}+\sin \theta c_{a'y\uparrow}\right)\left(-\sin \theta c_{a'x\downarrow}+\cos \theta c_{a'y\downarrow}\right)\nonumber\\&&-\left(\cos \theta c_{a'x\downarrow}+\sin \theta c_{a'y\downarrow}\right)\left(-\sin \theta c_{a'x\uparrow}+\cos \theta c_{a'y\uparrow}\right)\nonumber\\&&=\sum_{a}\cos 2\theta\left(c_{a'x\uparrow}c_{a'y\downarrow}- c_{a'x\downarrow}c_{a'y\uparrow}\right)-\sin 2\theta \left(c_{a'x\uparrow}c_{a'x\downarrow}- c_{a'y\uparrow}c_{a'y\downarrow}\right)\nonumber\\&&=\cos 2\theta \hat{\Delta}_{d_{xy}}-\sin 2\theta \hat{\Delta}_{d_{x^2-y^2}}
\end{eqnarray}
In addition, it is obviously inversion even. Thus this pair operator has the $d_{xy}$ symmetry.

\subsubsection{2. $d_{x^2-y^2}$-wave}
The $d_{x^2-y^2}$ pair operator is given by
\begin{equation}
\hat{\Delta}_{d_{x^2-y^2}}\equiv \sum_{a\mu}c_{a\mu\uparrow}c_{a\mu\downarrow}(-1)^{\mu}=\sum_{a}\left(c_{ax\uparrow}c_{ax\downarrow}-c_{ay\uparrow}c_{ay\downarrow}\right).\label{intra_s}
\end{equation}
When acted upon by $\hat{R}_{x,y}$ and $\hat{R}(\theta)$, it transforms as
\begin{eqnarray}
\hat{R}_{x}&:&\hat{\Delta}_{d_{x^2-y^2}}\to \sum_{a\mu}(-1)^{\mu+\mu}(-1)^{0+1}c_{a\mu\downarrow}c_{a\mu\uparrow}(-1)^{\mu}=\sum_{a\mu}c_{a\mu\uparrow}c_{a\mu\downarrow}(-1)^{\mu}=\hat{\Delta}_{d_{x^2-y^2}}\nonumber\\
\hat{R}_{y}&:&\hat{\Delta}_{d_{x^2-y^2}}\to \sum_{a\mu}(-1)^{\mu+\mu+2}(-1)^{0+1}c_{\bar{a}\mu\downarrow}c_{\bar{a}\mu\uparrow}(-1)^{\mu}=\sum_{a\mu}c_{\bar{a}\mu\uparrow}c_{\bar{a}\mu\downarrow}(-1)^{\mu}=\hat{\Delta}_{d_{x^2-y^2}}\nonumber\\
\hat{R}(\theta)&:&\hat{\Delta}_{d_{x^2-y^2}}\to \sum_{a}\left(\cos \theta c_{a'x\uparrow}+\sin \theta c_{a'y\uparrow}\right)\left(\cos \theta c_{a'x\downarrow}+\sin \theta c_{a'y\downarrow}\right)\nonumber\\&&-\left(-\sin \theta c_{a'x\uparrow}+\cos \theta c_{a'y\uparrow}\right)\left(-\sin \theta c_{a'x\downarrow}+\cos \theta c_{a'y\downarrow}\right)\nonumber\\&&=\sum_{a}\cos 2\theta \left(c_{a'x\uparrow}c_{a'x\downarrow}-c_{a'y\uparrow}c_{a'y\downarrow}\right)+\sin 2\theta\left(c_{a'x\uparrow}c_{a'y\downarrow}-c_{a'x\downarrow}c_{a'y\uparrow}\right)\nonumber\\
&&=\sin 2\theta \hat{\Delta}_{d_{xy}}+\cos 2\theta \hat{\Delta}_{d_{x^2-y^2}}.
\end{eqnarray}
In addition, it is obviously inversion even. Thus this pair operator has the $d_{xy}$ symmetry.
From the above transformation properties, it is apparent that the $d_{xy}$ and $d_{x^2-y^2}$ pair operators form a 2D representation of the point-group.

\subsection{c. f-wave}
There can be two types of f-wave (with total $S_z=0$ ): one is inter-orbital and the other is intra-orbital, let's analyze them separately.
\subsubsection{1. inter-orbital f-wave}
The pair operator of the inter-orbital f-wave is given by
\begin{equation}
\hat{\Delta}^{inter}_{f}\equiv \sum_{a}\left(c_{ax\uparrow}c_{ay\downarrow}+ c_{ax\downarrow}c_{ay\uparrow}\right)\epsilon_{a},\label{inter_f}
\end{equation}
where the sign $\epsilon_{a}$ is equal to 1(-1) for sublattice-A(B), respectively.
When acted upon by $\hat{R}_{x,y}$ and $\hat{R}(\theta)$, it transforms as
\begin{eqnarray}
\hat{R}_{x}&:&\hat{\Delta}^{inter}_{f}\to (-1)^{0+1}(-1)^{0+1}\sum_{a}\left(c_{ax\downarrow}c_{ay\uparrow}+ c_{ax\uparrow}c_{ay\downarrow}\right)\epsilon_{a}=\hat{\Delta}^{inter}_{f}\nonumber\\
\hat{R}_{y}&:&\hat{\Delta}^{inter}_{f}\to (-1)^{1+0}(-1)^{0+1}\sum_{a}\left(c_{\bar{a}x\downarrow}c_{\bar{a}y\uparrow}+ c_{\bar{a}x\uparrow}c_{\bar{a}y\downarrow}\right)\epsilon_{a}=-\hat{\Delta}^{inter}_{f}\nonumber\\
\hat{R}(\theta)&:&\hat{\Delta}^{inter}_{f}\to \sum_{a}\epsilon_{a}\left[\left(\cos \theta c_{a'x\uparrow}+\sin \theta c_{a'y\uparrow}\right)\left(-\sin \theta c_{a'x\downarrow}+\cos \theta c_{a'y\downarrow}\right)\right.\nonumber\\&&\left.+\left(\cos \theta c_{a'x\downarrow}+\sin \theta c_{a'y\downarrow}\right)\left(-\sin \theta c_{a'x\uparrow}+\cos \theta c_{a'y\uparrow}\right)\right]\nonumber\\&&=\sum_{a}\epsilon_{a}\left(c_{a'x\uparrow}c_{a'y\downarrow}+ c_{a'x\downarrow}c_{a'y\uparrow}\right)\nonumber\\&&=\left\{\begin{array}{cc}
{-\hat{\Delta}^{inter}_{f},\theta=\pm\pi/3}\\{\hat{\Delta}^{inter}_{f},\theta=\pm2\pi/3}\end{array}\right.
\end{eqnarray}
This pair operator is obviously inversion odd since it changes sign on the two sublattices. Thus, this pairing operator has f-wave symmetry.

\subsubsection{2. intra-orbital f-wave}
The intra-orbital f-wave pair operator is given by
\begin{equation}
\hat{\Delta}^{intra}_{f}\equiv \sum_{a\mu}c_{a\mu\uparrow}c_{a\mu\downarrow}\epsilon_{a}.\label{intra_s}
\end{equation}
When acted upon by $\hat{R}_{x,y}$ and $\hat{R}(\theta)$, it transforms as
\begin{eqnarray}
\hat{R}_{x}&:&\hat{\Delta}^{intra}_{f}\to \sum_{a\mu}(-1)^{\mu+\mu}(-1)^{0+1}c_{a\mu\downarrow}c_{a\mu\uparrow}\epsilon_{a}=\sum_{a\mu}c_{a\mu\uparrow}c_{a\mu\downarrow}\epsilon_{a}=\hat{\Delta}^{intra}_{f}\nonumber\\
\hat{R}_{y}&:&\hat{\Delta}^{intra}_{f}\to \sum_{a\mu}(-1)^{\mu+\mu+2}(-1)^{0+1}c_{\bar{a}\mu\downarrow}c_{\bar{a}\mu\uparrow}\epsilon_{a}=\sum_{a\mu}c_{\bar{a}\mu\uparrow}c_{\bar{a}\mu\downarrow}\epsilon_{a}=-\hat{\Delta}^{intra}_{f}\nonumber\\
\hat{R}(\theta)&:&\hat{\Delta}^{intra}_{f}\to \sum_{a}\epsilon_{a}\left[\left(\cos \theta c_{a'x\uparrow}+\sin \theta c_{a'y\uparrow}\right)\left(\cos \theta c_{a'x\downarrow}+\sin \theta c_{a'y\downarrow}\right)\right.\nonumber\\&&\left.+\left(-\sin \theta c_{a'x\uparrow}+\cos \theta c_{a'y\uparrow}\right)\left(-\sin \theta c_{a'x\downarrow}+\cos \theta c_{a'y\downarrow}\right)\right]\nonumber\\&&=\sum_{a}\epsilon_{a}\left(c_{a'x\uparrow}c_{a'x\downarrow}+ c_{a'y\uparrow}c_{a'y\downarrow}\right)
\nonumber\\&&=\left\{\begin{array}{cc}
{-\hat{\Delta}^{intra}_{f},\theta=\pm\pi/3}\\{\hat{\Delta}^{intra}_{f},\theta=\pm2\pi/3}\end{array}\right.
\end{eqnarray}
In addition, this pair operator is obviously inversion odd. Thus, it has the f-wave symmetry.

Since the inter and intra-orbital f-wave pair operators transform identically under all point operations, they are allowed to mix as
\begin{equation}
\hat{\Delta}_{f}\equiv \sum_{a}\epsilon_{a}\left(c_{ax\uparrow}c_{ay\downarrow}+ c_{ax\downarrow}c_{ay\uparrow}\right)+\beta\sum_{a\mu}\epsilon_{a}c_{a\mu\uparrow}c_{a\mu\downarrow}
\end{equation}

\subsection{d. p-wave}
There are two types of p-wave pair operators with the total $S_z=0$, i.e. $p_x$ and $p_y$, which form a 2D representation of the point-group.

The pair operators for $p_x$ and $p_y$ can be obtained from those of $d_{x^2-y^2}$ and $d_{xy}$ by adding the extra sign factor $\epsilon_{a}$, which are given by
 \begin{eqnarray}
\hat{\Delta}_{p_x}&\equiv& \sum_{a\mu}c_{a\mu\uparrow}c_{a\mu\downarrow}\epsilon_{a}(-1)^{\mu}=
\sum_{a}\epsilon_{a}\left(c_{ax\uparrow}c_{ax\downarrow}-c_{ay\uparrow}c_{ay\downarrow}\right)\nonumber\\
\hat{\Delta}_{p_y}&\equiv& \sum_{a}\epsilon_{a}\left(c_{ax\uparrow}c_{ay\downarrow}-c_{ax\downarrow}c_{ay\uparrow}\right).
\label{p}
\end{eqnarray}
 When acted upon by the elements of the point-group, they transform as
\begin{eqnarray}
\hat{R}_{x}&:&\hat{\Delta}_{p_{x}}\to \hat{\Delta}_{p_{x}}\nonumber\\
\hat{R}_{y}&:&\hat{\Delta}_{p_{x}}\to -\hat{\Delta}_{p_{x}}\nonumber\\
\hat{R}_{x}&:&\hat{\Delta}_{p_{y}}\to -\hat{\Delta}_{p_{y}}\nonumber\\
\hat{R}_{y}&:&\hat{\Delta}_{p_{y}}\to \hat{\Delta}_{p_{y}}\nonumber\\
\hat{R}(\theta)&:&\hat{\Delta}_{p_{x}}\to e^{3i\theta}\left(\cos 2\theta \hat{\Delta}_{p_{x}}+\sin 2\theta \hat{\Delta}_{p_{y}}\right)\nonumber\\
\hat{R}(\theta)&:&\hat{\Delta}_{p_{y}}\to e^{3i\theta}\left(-\sin 2\theta \hat{\Delta}_{p_{x}}+\cos 2\theta \hat{\Delta}_{p_{y}}\right).
\label{p_trans}
\end{eqnarray}
In addition, these pair operators are obviously inversion odd. Thus they have the $p_x$ and $p_y$ symmetries.

\subsection{e. p+ip'}
Although having the same symmetry, the $\left(p\pm ip'\right)_{\uparrow\uparrow,\downarrow\downarrow}$ pair operator discussed here is not the supposition of the $p_x$ and $p_y$ operators discussed in the previous section (the latter turns out not to be the leading unstable pair operator). Rather it is given by
\begin{equation}
\hat{\Delta}_{p+ip',\sigma}\equiv \sum_{a}\epsilon_{a}c_{ax\sigma}c_{ay\sigma}
\end{equation}
Due to the $\epsilon_{a}$ factor, this pair operator is inversion odd. Under $\hat{R}_{x,y}$ reflections, the $\hat{\Delta}_{p+ip',\sigma}$ for opposite spins transform into each other (since reflection change the sign of the spin).

In the main text, we show in Eq.(\ref{conjg}) that the Cooper scattering amplitude for $\uparrow\uparrow$ is the complex conjugate of that for $\downarrow\downarrow$. Therefore, we can choose a gauge under which the expectation value of $\hat{\Delta}_{p+ip',\uparrow}$ is the complex conjugate of that of $\hat{\Delta}_{p+ip',\downarrow}$. Under this gauge, we define
 \begin{eqnarray}
\hat{\Delta}^{R}_{p+ip'}&\equiv& \hat{\Delta}_{p+ip',\uparrow}+\hat{\Delta}_{p+ip',\downarrow}\nonumber\\
\hat{\Delta}^{I}_{p+ip'}&\equiv& i\left(\hat{\Delta}_{p+ip',\uparrow}-\hat{\Delta}_{p+ip',\downarrow}\right).
\end{eqnarray}
which are the real (imaginary) part of the $p+ip'\downarrow\downarrow$ pair operator. It is straightforward to show that
\begin{eqnarray}
\hat{R}_{x}&:&\hat{\Delta}^{I}_{p+ip'}\to\hat{\Delta}^{I}_{p+ip'}\nonumber\\
\hat{R}_{y}&:&\hat{\Delta}^{I}_{p+ip'}\to-\hat{\Delta}^{I}_{p+ip'}\nonumber\\
\hat{R}_{x}&:&\hat{\Delta}^{R}_{p+ip'}\to-\hat{\Delta}^{R}_{p+ip'}\nonumber\\
\hat{R}_{y}&:&\hat{\Delta}^{R}_{p+ip'}\to \hat{\Delta}^{R}_{p+ip'}\nonumber\\
\hat{R}_{\theta}&:&\hat{\Delta}^{I}\to e^{3i\theta}(\cos 2\theta\hat{\Delta}^{I}+\sin 2\theta \hat{\Delta}^{R})\nonumber\\
\hat{R}_{\theta}&:&\hat{\Delta}^{R}\to e^{3i\theta}(-\sin 2\theta\hat{\Delta}^{I}+\cos 2\theta\hat{\Delta}^{R}).\label{pip'}
\end{eqnarray}
Comparing Eq.(\ref{pip'}) and Eq.(\ref{p_trans}), we find that $\hat{\Delta}^{I}$ and $\hat{\Delta}^{R}$ transform identically as $\Delta_{p_x}$ and $\Delta_{p_y}$. Therefore, we have obtained the $p_y+ip_x\uparrow\uparrow;p_y-ip_x\downarrow\downarrow$ pair operators, i.e $(p\pm ip')_{\uparrow\uparrow,\downarrow\downarrow}$.

\end{widetext}

\end{document}